\documentstyle[12pt]{article}
\input epsf.tex

\begin{document}
\bigskip
\hskip 3.7in\vbox{\baselineskip12pt
\hbox{NSF-ITP-96-60}\hbox{hep-th/9607050}}
\bigskip\bigskip\bigskip\bigskip

\centerline{\Large \bf String Duality}
\medskip
\centerline{\bf A Colloquium}

\bigskip\bigskip
\bigskip\bigskip

\centerline{\bf Joseph Polchinski} 
\medskip
\centerline{Institute for Theoretical Physics}
\centerline{University of California}
\centerline{Santa Barbara, CA\ \ 93106-4030}
\centerline{e-mail: joep@itp.ucsb.edu}
\bigskip\bigskip

\begin{abstract}
\baselineskip=16pt
The strong coupling limit of a quantum system is in general quite
complicated, but in some cases a great simplification occurs: the
strongly coupled limit is equivalent to the weakly coupled limit of
some other system.  In string theory conjectures of this type go back
several years, but only in the past year and a half has it
been understood to be a general principle applying to
all string theories.  This has improved our understanding of string
dynamics, including quantum gravity, in many new and sometimes
surprising ways.  I describe these developments and put them in the
context of the search for the unified theory of particle physics and
gravity.
\end{abstract}
\newpage
\baselineskip=18pt

\section{Introduction}

String duality is a recently discovered symmetry of string theory. 
String theory itself is over twenty-five years old, and has been under
intensive development since 1984 as the leading candidate for a
unified theory of particle physics and gravity.  The reason that some
of its symmetry has been overlooked until now is simple, and
important: string duality is not manifest in the weak-coupling
perturbation expansion by which the theory is usually studied, but it
is a property of the exact theory.  As a result, string duality gives
information about the behavior of string theory at strong coupling. 
In a period of a little over a year, we have gone from near-complete
ignorance of the behavior of strongly-coupled strings to a rather
detailed understanding of the intricate dynamics which occurs, at
least in vacua having enough supersymmetry, and the subject continues
to develop at a rapid pace.

The central idea of string duality is that the strongly coupled limit
of any string theory is equivalent to the weakly coupled limit of some
other theory.  All string theories are connected in this way, as well
as something new and surprising: an eleven-dimensional theory known
provisionally as `M-theory.'  Besides the ordinary vibrating strings
which are the basic quanta of string theory, the multiplets of string
duality include smooth classical objects (solitons), singular
classical objects (black holes), and a new type of topological defect
which is unique to string theory (D-branes).  With the
improved understanding of string dynamics it has become possible to
address one of the long-standing problems of quantum gravity---to
count the number of states of certain black holes in a controlled way,
giving for the first time a statistical mechanical interpretation to
the Bekenstein-Hawking entropy.

Beyond these specific results, string duality has greatly changed the
way string theorists think about the fundamental principles of the
theory.  Many ideas which once seemed to be central are now seen as
technicalities, while other ideas which were neglected are now
central.  In this Colloquium I would like to try to explain these
developments and to put them in the context of the search for the
unified theory of particle physics and gravity.

\section{String Theory: A Review}

\subsection{Strings as a Unified Theory}

I will use units in which $\hbar = c = 1$.  The gravitational coupling
$G_{\rm N}$ is then a length-squared, 
\begin{equation} 
G_{\rm N} = l_{\rm P}^2 
\end{equation}
where $l_{\rm P}$ is the Planck length, $1.6 \times 10^{-33}$ cm. 
This is the natural length scale for the effects of quantum gravity to
become important and so for the unification of gravity with the other
interactions. It is far shorter than the length scales which can be
probed directly; the corresponding energy scale is  
\begin{equation}
M_{\rm P} = l_{\rm P}^{-1} = 1.2 \times 10^{19}\ {\rm GeV},
\end{equation} 
far beyond the reach of accelerators.  Thus, to
construct a unified theory one must rely heavily on theoretical
reasoning such as the internal consistency of the theory, and any
experimental tests will be indirect. 

Fortunately, consistency is a very restrictive guide.  General
relativity is a nonrenormalizable field theory, meaning that its
quantum mechanical perturbation theory has uncontrollable
divergences.  To see the significance of this, let us recall the 
four-fermi theory of the weak interaction.\footnote
{A review of weak interaction theory can be found in Commins and
Bucksbaum (1983).}
The weak
interaction was originally described as an interaction of four
fermionic fields at a spacetime point as depicted in figure~1a.
\begin{figure}
\begin{center}
\leavevmode
\epsfbox{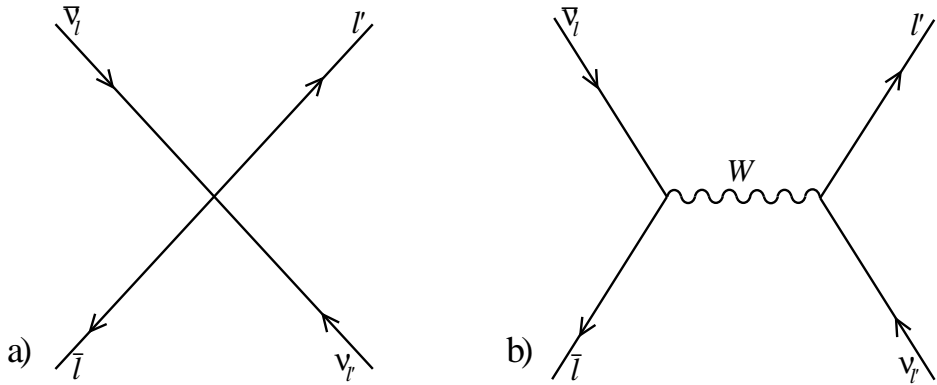}
\caption{a) Leptonic weak interaction in four-fermi theory.
b) The same interaction in Weinberg-Salam theory.  At short distance
the contact interaction is resolved into the exchange of a $W$ boson.}
\end{center}
\end{figure}
The weak coupling constant $G_{\rm F}$ also has units of
length-squared, or inverse energy-squared.  In a process with a
characteristic energy $E$ the effective dimensionless coupling is
then $G_{\rm F} E^2$.  It follows that at sufficiently high energy the
coupling becomes arbitrarily strong, and this also implies divergences
in the perturbation theory.  A second order weak amplitude is
dimensionally of the form 
\begin{equation} G_{\rm F}^2 \int^\infty E' dE', 
\end{equation} 
where $E'$ is the energy of the virtual state in
the second order process, and this diverges at large energy.  In
position space the divergence comes when the two weak interactions
occur at the same spacetime point (high energy = short distance).  The
divergences become worse at each higher order of perturbation theory
and so cannot be controlled even with renormalization.

The natural interpretation of such divergences is that the theory one
is working with is only valid up to some energy scale, beyond which
new physics appears.  The new physics should have the effect of
smearing out the interaction in spacetime and so soften the high energy
behavior.  One might imagine that this could be done in many ways, but
in fact the combined constraints of Lorentz invariance and causality
are very restrictive.  This is because Lorentz invariance requires
that if the interaction is spread out in space it is also spread out
in time.  In fact, for the weak interaction there is only one known
way to solve the short-distance problem.  This is depicted in
figure~1b, where the four-fermi interaction is resolved into the
exchange of a vector boson.  Moreover, this vector boson must be of a
very specific kind, coming from a spontaneously broken gauge
invariance.  This is the only known solution, and in fact it is the
one that nature chooses.\footnote{It could also have been that the
divergences are an artifact of perturbation theory but do not appear
in the exact amplitudes.  This is a logical possibility, a `nontrivial
fixed point.'  Although conceivable, it seems unlikely, and it is not
what happens in
the case of the weak interaction.}

For gravity the discussion is much the same.  The gravitational
interaction is depicted in figure~2a.
\begin{figure}
\begin{center}
\leavevmode
\epsfbox{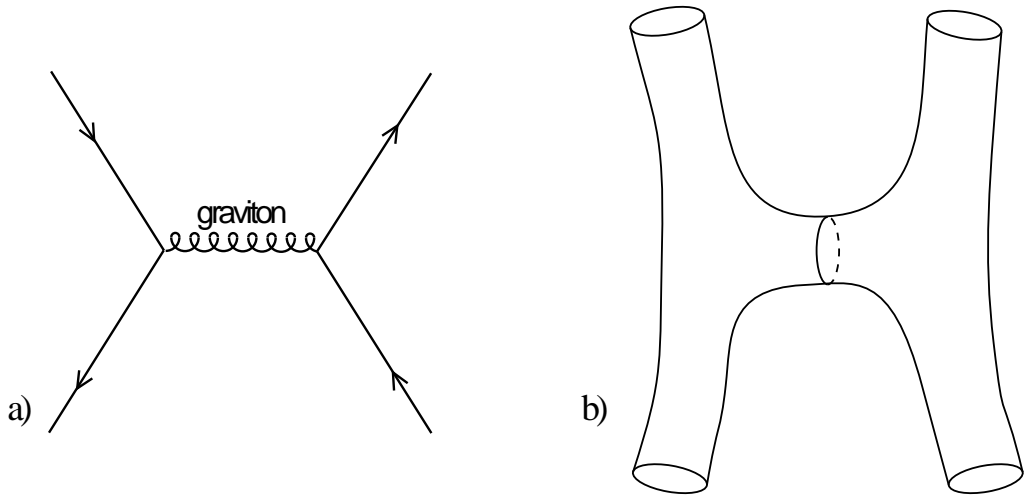}
\caption{a) Exchange of a graviton between two elementary particles.
b) The same interaction in string theory.  The amplitude is given by
the sum over histories, over all embeddings of the string
world-sheet in spacetime.  The world-sheet is smooth:
there is no distinguished point at which the interaction occurs (the
cross section on the intermediate line is only for illustration).}
\end{center}
\end{figure}
As we have noted already, the
gravitational  coupling $G_{\rm N}$ has units of length-squared and so
the dimensionless coupling is $G_{\rm N} E^2$.  This grows large at
high energy and signifies a nonrenormalizable perturbation
theory.\footnote{Note that the bad gravitational interaction of
figure~2a is the same graph as the smeared-out weak interaction of
figure~1b.  However, its high energy behavior is worse because gravity
couples to energy rather than charge.}  Again the natural suspicion is
that short-distance physics smears out the interaction, and again
there is only one known way to do this, shown in figure~2b.  It
involves a bigger step than in the case of the weak interaction: it
requires that at the Planck length the graviton and other particles
turn out to be not points but one-dimensional objects, loops of
`string.'\footnote{Reprints of early papers can be found in Schwarz
(1985). Green, Schwarz, and Witten (1987) is a text with extensive
references, including papers from the first period of string theory
(roughly 1969-1974) when strings were studied as a possible theory of the
strong interaction.}

It is certainly not obvious that this should be the right thing to do,
but if one tries to make a consistent Lorentz-invariant quantum theory
of one-dimensional objects, one finds that it is possible but that the
theory is highly constrained.  In particular, one of the states of the
string is massless and has spin two.  Consistency
requires that such a particle be a graviton, that its long-wavelength
interactions be described by general relativity, and indeed this is
what is found from the sum-over-histories depicted in figure
2b~(Yoneya, 1974; Scherk and Schwarz, 1974, 1975).  In a sense string
theory predicts gravity, in that every consistent string theory
includes general relativity, even though the theory initially seems to
be formulated in flat space.  Further, calculation of the quantum
corrections to amplitudes shows that they are finite.

It is also not obvious why {\it only} this idea should work.  Many
others have been tried without success.  One natural question is,
why one-dimensional objects and not two-dimensional or higher?  For
this at least there is a simple answer.  Extended objects have an
infinite number of internal degrees of freedom (the Fourier modes
describing their shape).  Spreading out point particles into extended
objects softens the spacetime divergences but introduces potential new
divergences from the internal degrees of freedom.  The latter grow
worse as the dimension of the object increases, and the
one-dimensional case is the only one for which both the spacetime and
internal behavior is under control.

It could be that we suffer from a failure of imagination, and that
there are other solutions.  However, we see from the example of the
weak interaction that if we can find even one solution to a
short-distance problem we should take it very seriously and see where
it leads.  The result here is quite striking.  Besides the graviton,
other states of the string with different internal oscillators excited
behave as gauge bosons (Neveu and Scherk, 1972), and yet others are
fermions (Ramond, 1971).  In fact, string theory automatically
incorporates (and in some ways generalizes) three earlier ideas for
explaining the patterns in the Standard Model, namely grand
unification, supersymmetry (Gliozzi, Scherk and Olive, 1977), and
Kaluza-Klein theory.  Some of the simplest string theories, the
Calabi-Yau models, closely resemble unified versions of the
supersymmetric Standard Model.\footnote{Three key papers that
established this were Green and Schwarz (1984), Gross, Harvey,
Martinec and Rohm (1985), and Candelas, Horowitz, Strominger, and
Witten (1985).  Green, Schwarz and Witten (1987) gives extensive
coverage to these developments.}

As a particle physicist I usually emphasize these particle physics
motivations for string theory, but many of those educated in general
relativity and in mathematics find string theory compelling for reasons
that are not entirely the same.  From Newton's gravity to
Einstein's, from quantum mechanics to non-Abelian gauge theory, new
physical theories have often required new mathematics, or least
mathematics that had not previously been used in physics.  If one
searches for higher symmetries or
other more mathematical structures that might be useful in physics,
one finds many connections to string theory.  One reason I emphasize
this now is that in the recent work on string duality, all of these
different points of view have had a role to play.

It is worthwhile to note that these three kinds of
motivation---solving the divergence problem, the connection with
geometry, and explaining the broad patterns in the Standard Model,
were also present in the weak interaction.  Weinberg (1980) emphasized
the divergence problem as I have done.  Salam (1980) was more guided by
the idea that non-Abelian gauge theory was a beautiful structure that
should be incorporated in physics.  Experiment gave no direct
indication that the weak interaction was anything but the pointlike
interaction of figure~1a, and no direct clue as to the new physics
that smears it out, just as today it gives no direct indication of
what lies beyond the Standard Model.  But it did show certain broad
patterns---universality and the $V-A$ structure, which were telltale
signs of a gauge interaction.  It appears that nature is kind to us,
in providing many trails to a correct theory.

\subsection{String Theory before Duality}

The key question is, how do we go from explaining broad patterns to
making precise predictions?  To understand the situation, it is useful
to look again to history, to the state of quantum field theory in the
early '60's.  At that time there was a good technical control of the
weak-coupling perturbation theory, the Feynman graph expansion, but
little else.  Important dynamical ideas were missing, such as the
Higgs mechanism, dynamical symmetry breaking, and confinement.  These
have largely to do with the fact that the vacuum, the ground state of
quantum field theory, is generally a more interesting and less
symmetric object than one might naively expect from the Hamiltonian. 
Without understanding these ideas one cannot make sense of
the Standard Model: it would seem to predict fractionally charged
particles and a long-ranged force coupled to isospin.

Beyond these dynamical ideas, it remained to discover the defining
principle of the theory---that one should organize the physics not
{\it graph by graph} but rather {\it length scale by length scale}
with the shortest distances first.\footnote {Wilson (1983) gives a
fascinating account of this discovery.} This realization not only made
it possible to define the theory beyond perturbation theory, but also
provided a framework for understanding the dynamics, both analytically
and numerically.

The situation is similar in string theory.  There has been a good
technical control of the perturbation theory, but little else, at
least up until the recent developments.  The need to understand the
dynamics is even more acute than in the case of the Standard Model for
two reasons.  The first is that because the Planck energy is so large,
there is no hope to explore a `partonic' regime where the stringy
behavior is directly visible---rather, we see only the extreme low
energy limit of string theory, filtered through all the dynamics at
intervening scales.  The second reason has to do with one of the very
attractive features of string theory, that it has no free
dimensionless parameters at all.  Instead it has, in perturbation
theory, many degenerate ground states parameterized by the values of
various scalar fields (moduli).  The parameters of the Standard Model
come ultimately from the values of these scalars, so it is necessary to
understand the dynamics which selects one of the many ground states.
String theory contains quantum field theory as its low
energy limit, so all of the familiar dynamics of the latter is still
present, but because string theory has many more degrees of freedom it
is likely to have interesting new dynamics of its own.

Beyond this, the central defining principle of string theory is not
known.  We are trying to answer the question `What is string
theory?', just as Wilson and others addressed the question `What is
field theory?'.  The various properties that make string theory an
attractive unifying idea also imply that the theory exists as
something more than an asymptotic weak-coupling expansion.  So there
is good reason to expect that we will find a central principle as
powerful as that in field theory, which will again enable us to better
understand the dynamics.

With this background I can summarize what has been learned from string
duality.  We have learned a number of new dynamical ideas though
certainly not yet all we need.  We have not yet found the central
principle but we have many new clues, some of which are surprising
and have taken us in unexpected directions.

\section{Some Ideas} 

String duality involves an interplay of many different ideas from
quantum field theory and string theory.  In this part I would like
to explain some of the central principles.

\subsection{Strong Coupling and Duality}

Here are two ways to think about the meaning of the
coupling constant $g$ of a quantum field theory.  The first is in terms
of the weak-coupling perturbation expansion.  The amplitude $A$ for any
process can be expanded  
\begin{equation}
A = \sum_{n=0}^\infty c_n g^n ,
\end{equation}
where $g$ is the amplitude for a single interaction to
occur.\footnote{For simplicity this is written for the case that
there is only one kind of interaction.  Incidentally, it will
frequently be the case that only even $n$, or only odd, contribute to
a given amplitude.}
The coefficient $c_n$ is given by the sum
over Feynman graphs with $n$ interaction vertices.

Interpreted with sufficient care this is an asymptotic series, meaning
that it gives an approximation of any desired accuracy by taking $g$
to be sufficiently small.  Thus it is also a good qualitative guide
to the small-$g$ physics.  But for $g$ of order $1$ or larger it is
of limited usefulness and can miss important aspects of the physics.

Another way to think about the meaning of $g$ is in terms of the
quantum fluctuation of the fields.  For the purpose of this
discussion it is useful to keep explicit factors of $\hbar$; we will
return to $\hbar=1$ units after section 3.2.
When $g$ is small the fluctuations are small and their
equations of motion can be approximated by linear equations---in
other words, small $g$ is approximately free field theory.  For larger
$g$ the fluctuations of the fields, and the nonlinearities, become
larger.\footnote{A more detailed way to see the connection between the
value of $g$ and the size of the quantum fluctuations is in terms of
the path integral, the sum over field histories weighted by
$e^{iS/\hbar}$ with $S$ the classical action.  The fields can be
rescaled in such a way that $g$ appears in the action only as an
overall factor $g^{-2}$, so $g$ and
$\hbar$ appear only in the combination $g^2 \hbar$.  It is then clear
that small $g$ and small $\hbar$ are equivalent with this scaling of
the fields.  To relate this to the familiar example of QED, one must
note that the electric charges of individual quanta are related to the
charge in the classical field action by $e = \hbar g$, so the familiar
expansion parameter $\alpha = e^2/4\pi\hbar$ is indeed $g^2
\hbar/4\pi$.} In QCD, for example, the coupling is very
strong at long distance, and correspondingly the color magnetic fields
undergo large fluctuations in the vacuum---this is the source of
confinement.

Weak/strong duality (in field theory or string theory) means that as 
the string coupling $g$ becomes large, one can find new `dual' fields
whose fluctuations become {\it small}---they are characterized by a
new coupling $g'$ which is something like $1/g$.  This is similar to
a Fourier transform, where a function which becomes spread out
in position space can become very narrow in momentum space.  Here
though, the Fourier transform is in a complicated nonlinear 
field space.

String theory includes gravity, and so one of the fields is the
spacetime metric.  One might therefore have expected strongly coupled
string theory to have new and very exotic physics, including large
fluctuations of the spacetime geometry and all other fields.  This
might correspond to some sort of confined phase of gravity.  One of
the surprises of string duality is that this is not so.  As $g \to
\infty$, $g' \to 0$ and so the metric of the dual theory behaves
more and more classically.

It is quite likely that in nature the string coupling is
close to~1 rather than to~0 or~$\infty$, so that string duality does
not allow one to relate the theory directly to a weakly coupled theory
in which one can calculate accurately.  But having an understanding
of both limits, large and small coupling, constrains the kinds of
qualitative physics that can occur at
intermediate coupling.  Even more important, duality gives a great
deal of information about the exact theory and its symmetries, 
information that is not contained in the perturbation expansion, and
so new clues to the answer to `What is string theory?'.

\subsection{Electric/Magnetic Duality} 

Another perspective on the meaning of duality starts with Maxwell's
equations,
\begin{eqnarray}
&&\mbox{\boldmath$\nabla$} \cdot {\bf E} = \rho\qquad
\mbox{\boldmath$\nabla$} \times {\bf E} + {\bf \dot B} = 0 \nonumber\\
&&\mbox{\boldmath$\nabla$} \cdot {\bf B} = 0\qquad
\mbox{\boldmath$\nabla$} \times {\bf B} - {\bf \dot E} = {\bf j}.
\end{eqnarray}
The symmetry between $\bf E$ and $\bf B$ in these equations is
striking.  If one ignores the sources, or adds magnetic sources,
the equations are invariant under $\bf E \to B$, $\bf B \to - E$.

This curious fact was made more interesting by Dirac's 
study (1931) of the quantum mechanics of a charge moving in a magnetic
monopole field.  He found that the wavefunction could be consistently
defined only if the electric charge $e$ and magnetic charge $q$
satisfy a quantization condition
\begin{equation}
eq = 2\pi\hbar n. \label{dirac}
\end{equation}
Note that if a monopole of some charge $q$ exists, then all
electric charges must be multiples of the unit $2\pi\hbar/q$.
This would `explain' why the magnitudes of the electron and
proton charges should be {\it exactly} equal, a fact known to hold to
one part in $10^{21}$.

The subject took another step forward when 't Hooft (1974) and
Polyakov (1974) showed that in any grand unified theory, magnetic
monopoles actually do exist.  They are classical solutions, with a
nonsingular core whose size is set by the scale of spontaneous
symmetry breaking.

At weak coupling, the electrically and magnetically charged objects
look very different.   The electrically charged objects are weakly
coupled and have pointlike interactions.  The magnetically charged
objects are strongly coupled---by the Dirac condition~(\ref{dirac}),
the magnetic fine structure constant $\alpha_m = q^2 /4\pi \hbar =
n^2/4\alpha$ is roughly the reciprocal of the usual one---and as
noted above they have cores of finite size.  

Montonen and Olive (1977) were led by various evidence to conjecture
that at strong coupling the situation would be reversed: the
electrically charged objects would be strongly coupled and have
nonsingular cores, while the magnetically charged objects would become
weakly coupled and pointlike.  The strongly coupled theory would be
equivalent to weakly coupled theory in which the basic quanta carried
magnetic rather than electric charges.  In subsequent work this
conjecture was refined (Witten and Olive, 1978; Osborn, 1979).  It was
argued to hold specifically in supersymmetric gauge theories, in
particular
$N=4$ theories ($N$ is the number of conserved supersymmetries).
Relating this discussion to the previous section, the weakly coupled
dual fields described there would be the fields corresponding to the
magnetic quanta.\footnote{Returning to QCD, the strongly fluctuating
color magnetic fields would be roughly dual to a state in which the 
color electric field goes rapidly to zero at long distance.  This is
indeed one way to understand confinement, though any sort of precise
quantitative duality in QCD is unlikely.}

These conjectures were greeted with wide
skepticism because the evidence for them was rather
circumstantial, while attempts to construct the dual
fields directly in terms of the original ones did not
succeed.\footnote{Dual fields can be constructed for some quantum
field theories in two spacetime dimensions, the Ising model and
the Sine-Gordon/Thirring models being the simplest examples.}
This skepticism is now largely gone, not because the dual fields have
been found (they have not), but because of a substantial
strengthening of the circumstantial evidence.  This evidence, based
on supersymmetry, will be described in the next section.

The reader should notice that this whole section has been concerned
with duality in field theory, not string theory.  The extension to
string theory will be discussed in section~3.5.
 
\subsection{Supersymmetry}

To test the duality conjectures we need to be able to say something 
about the physics of the strongly coupled theories.  The methods
available for this are very limited, but in supersymmetric theories it
is possible.

Supersymmetry is of interest for a number of reasons.  It is likely
that it is associated with the breaking of the electroweak symmetry. 
If so, it should be discovered by the LHC if not before, and it is the one
piece of new physics associated with string theory that might be
accessible at accelerators.  Beyond this, it is an appealing
mathematical structure that extends general coordinate invariance and
unifies fermions and bosons.  Any consistent string
theory must have supersymmetry, at least at the Planck scale.
Finally, supersymmetric field theories, and string theories with
unbroken supersymmetry below the Planck scale, have various nice
properties that make them easier to study.

It is difficult to give an intuitive picture of supersymmetry.  One
way to think about it is that in addition to the usual spacetime
dimensions, whose coordinates are real numbers $x^\mu$ (and so their
multiplication commutes), there are additional `fermionic' coordinates
$\theta_\alpha$ which anticommute, $\theta_\alpha\theta_\beta =
-\theta_\beta\theta_\alpha$.  These extra dimensions have no size;
the anticommuting property means that they are essentially
infinitesimal.  One can make sensible field theories on this space.
In fact, they can be thought of as quantum field theories on ordinary
spacetime but with fields that have both fermionic and bosonic
components, with a symmetry relating the masses and charges of the
particles with different statistics.

To see why supersymmetry is valuable in studying strong coupling we
need the algebra of the quantum mechanical generators of the
symmetry.  For an ordinary internal symmetry, such as baryon
number, the generator $G$ commutes with the Hamiltonian $H$,
\begin{equation}
[G,H] = 0.
\end{equation}
This is the definition of a symmetry.  For supersymmetry, with
generator $Q$, there is an additional relation (see, for example,
Haag, Lopuszansky, and Sohnius, 1975)
\begin{equation}
[Q,H] = 0 \qquad \{Q,Q\} = H + G. \label{susy}
\end{equation}
In the second, anticommutation, relation, both the Hamiltonian and
various internal generators appear on the right-hand side.  For
clarity we have omitted the indices which would distinguish the various
supersymmetry and  internal symmetry generators and various associated
constants, so this anticommutation relation is schematic.

The anticommutation relation has the Hamiltonian on the right-hand side, and
as a result supersymmetry gives much more information about the
dynamics than ordinary internal symmetries.  To see one example of
this (Witten and Olive, 1978), consider a one-particle state $| \psi
\rangle$ which has the special property that is invariant under part
of the supersymmetry algebra; in other words, $Q| \psi \rangle = 0$ for some
of the
$Q$'s.  This is known as a {\it Bogomol'nyi-Prasad-Sommerfeld (BPS)
state}.  Take the expectation
value of the anticommutator in eq.~(\ref{susy}),
\begin{equation}
\langle \psi | \{Q,Q\} | \psi\rangle\ =
\ \langle \psi | H | \psi\rangle\ + 
\langle \psi | G | \psi\rangle.
\end{equation}
For those $Q$'s which annihilate $| \psi\rangle$, the left hand side
is zero.  The two terms on the right are just the mass of the
particle and its $G$-charge.  Thus, the mass of any BPS particle 
is determined entirely by its charge.  This is a consequence of
symmetry and does not depend on dynamics at all; in particular it
remains true even if the coupling is large.

Further analysis puts strong constraints on the interactions and on
the phase diagram.  For example, any BPS state
with zero charge has zero energy.  The anticommutation relations
imply that no other state can have lower energy, so any such state
will be the ground state.  A typical supersymmetric theory has many
such states, which are characterized by the expectation values of some
scalar fields.  All these states must be degenerate.  This is
similar to spontaneous symmetry breaking, but with spontaneous
symmetry breaking the degenerate vacua all have the same physics
while here they are physically inequivalent---they are not related to
one another directly by any symmetry, rather the degeneracy follows
indirectly from the BPS argument.  The scalar fields which label
these vacua, known as moduli, must be massless for the same reason
that Goldstone bosons are massless.\footnote
{In realistic theories the degeneracy is removed and the moduli made
massive by supersymmetry breaking.}
The low energy physics,
the phase structure, is determined by the physics of the
moduli, which is strongly constrained by supersymmetry.

Applying these methods to test the $N=4$ duality conjecture, one
first finds that the BPS mass formula is duality symmetric.\footnote
{Recent reviews of this subject are Sen (1994) and Harvey (1996).}
That is,
electrically charged BPS states at coupling $g$ have the same masses
as magnetically charged states at coupling $1/g$.  Secondly, one can
compare the degeneracies of BPS states with different charges and
spins.  The degeneracy of magnetic monopoles can be determined by
semiclassical methods when $g$ is small.  In $N=4$ theories there
cannot be a phase transition as the coupling is varied and the
supersymmetry algebra prevents the number of BPS states from changing,
so this also determines the degeneracy at strong coupling.
For
$N=4$ theories it is the same as that of the electrically charged
states in the dual theory.  As another test, the effective low energy
physics of the moduli is duality invariant.

This evidence was widely regarded as unconvincing, an
accidental consequence of supersymmetry, until
Seiberg (1994) and others began to apply these methods
systematically to determine the phase structures of theories with less
supersymmetry, $N=1$ and $N=2$. The arguments in these cases are more
intricate, the physics is richer, and there are more consistency
checks.  Many more examples of duality were found (reviewed in
Seiberg, 1995, and Intriligator and Seiberg, 1995), as well as
convincing evidence for various associated phenomena such as composite
gauge bosons.\footnote{In a few cases, like $N=4$, the duality holds
at all energies.  In many others it is holds only in the low energy
theory.}  Duality in supersymmetric gauge theory is now
well-established.  It is important to note, though, that the dual
fields still have not been constructed.  It is believed that to make
duality manifest one may need a new formulation of these quantum field
theories, something more intrinsically quantum-mechanical than the
path integral over classical histories.  It is also suspected that an
understanding of the dualities in field theory will ultimately come
from string theory.

\subsection{Higher Dimensions}

We will need one more idea in order to fully understand
string duality.  This is that spacetime may have more than four
dimensions.  The three spatial dimensions we see are expanding, and
once were highly curved.  It is a logical possibility that there
are other dimensions which did not expand but remain small and
highly curved.  Moreover, this is an attractive idea for a number of
reasons.

A good model for physics in such a spacetime is a waveguide, a cavity
which has some finite cross section in the $x$-$y$ plane and is very
long in the $z$ dimension.  Seen from far away, or with low
resolution, this looks one-dimensional.  An infinite number of
different fields (functions of $z$ and $t$) with different dispersion
relations (masses) move along the waveguide.  Seen up close, the
three-dimensional structure is evident and one sees that there is only one
field, the electromagnetic field, and that the different dispersion relations
come from modes with different $x$, $y$ dependence.

Physics is much the same in a spacetime with three large spatial
dimensions and additional ones which are small and compact.  In
nature, the additional dimensions would be quite small, close to the
Planck scale, so we would have seen only the very lowest modes.
One reason this is attractive is that it unifies gravitational and
gauge interactions.  Depending on whether its polarization is aligned
along the long or compact directions, a higher-dimensional graviton
can look like a graviton, photon, or scalar from the
lower-dimensional point of view.  This is the Kaluza-Klein mechanism.
In addition, the Dirac equation on such a space typically gives rise
to multiple copies of the same set of quantum numbers (that is,
generations) in the lower-dimensional spectrum.

Consistent weakly-coupled string theories necessarily live in ten
spacetime dimensions.  The origin of this condition is difficult to
explain in simple terms, but it can be understood in various ways. 
Calculation of quantum effects shows that they spoil essential
symmetries unless the dimension is ten.  Also, the properties of
fermions depend in an essential way on the number of dimensions and
ten is special here for a number of reasons related to supersymmetry. 
For example, ten is the highest dimension in which the number of
states of a massless vector (eight: the spacetime dimension minus the
timelike and longitudinal polarizations) can be equal to the number of
states of a massless fermion---in higher dimensions the spinor
representations are too large.

The field equations of ten-dimensional string theory are consistent
with four-dimensional physics that looks very much like what we
have.  Some of the simplest solutions give rise to the same gauge
groups and matter representations found in grand unification
(Candelas, Horowitz, Strominger, and Witten, 1985).  It is notable
that the four-dimensional gauge couplings can be chiral, meaning that
the right- and left-handed fermions have different gauge
interactions.  This was impossible for a number of previously
considered unifying ideas including standard Kaluza-Klein theory
(Witten, 1981).

For future reference it is useful to look at a simple example.  This
is a massless scalar field in five spacetime dimensions, with one
dimension periodic with period $2\pi R$.  Let $x^\mu$ with $\mu =
0,1,2,3$ be the coordinates for the large dimensions, and $x^4$ be
the periodic coordinate; the index $M$ runs over all five values.  The
wave equation is 
\begin{equation}
0 = \eta^{MN} \frac{\partial^2}{\partial X^M \partial X^N} \phi(x)
= \left(
\eta^{\mu\nu} \frac{\partial^2}{\partial X^\mu \partial X^\nu}
+ \frac{\partial^2}{\partial^2 X^4 }\right)\phi(x) .
\end{equation}
As in the waveguide, separate variables and expand the $x^4$
dependence in a complete set of modes,
\begin{equation}
\phi_n(x) = \phi_n(x^\mu) e^{inx^4/R}.
\end{equation}
Then the $n$'th mode satisfies
\begin{equation}
0 
= \left(\eta^{\mu\nu} \frac{\partial^2}{\partial X^\mu \partial X^\nu}
- \frac{n^2}{R^2} \right) \phi_n(x) 
\end{equation}
which is the Klein-Gordon equation for a scalar field of mass $M =
n/R$. At low energies only the $n=0$ mode is detectable, but at
energies above $1/R$ one sees a characteristic infinite tower of
states.  For this simple geometry the states are spaced evenly
in mass.  More complicated compact spaces would of course give a more
complicated spectrum, but the average density of states depends only
on the number of additional dimensions.  From another point of view,
consider how the spectrum behaves as $R \to \infty$: the infinite
tower comes down in mass and forms the continuum characteristic of a
noncompact dimension.

\subsection{String Duality} 

String theory includes gauge theory.  Weak/strong duality in gauge
theory is thus necessary if the same is to be true in string theory, but
it is not sufficient.  After the work of Seiberg, it was still
possible that duality was a property only of the low energy limit of
string theory, not of the full massive spectrum.

There were a number of early duality conjectures in string theory.
A partial list includes the self-duality of $N=1$ heterotic string
theories in four\footnote
{Four is the number of large dimensions, meaning that six are
compactified.} dimensions (Font, Ibanez, Lust, and Quevedo, 1990),
self-duality of $N=4$ heterotic strings in four dimensions
(Font, Ibanez, Lust, and Quevedo, 1990; Sen, 1994), self-duality of
the heterotic string in six dimensions (Duff,1995), duality of certain
string theories with theories of five-dimensional objects
(five-branes) in ten dimensions (Duff, 1988; Strominger, 1990), and a
relation---not specifically duality---of other string theories in ten
dimensions to theories of membranes (two-branes) in eleven dimensions
(Duff, Howe, Inami, and
Stelle, 1987).  These conjectures did not attract broad attention, due to
their rather scattered nature and the limited
evidence for any of them.

The subject took a major step forward with the papers of Hull and Townsend 
(1995), Townsend (1995), and Witten (1995a).  These authors proposed a
nearly complete set of duals for all known string theories with at least
$N=4$ supersymmetry. 
These new proposals elevated duality from a set of isolated conjectures to a
general principle applying to any string theory in any dimension.  The
structure is quite rigid---for each string theory the spectrum and symmetries
determine a unique candidate for its dual, and the entire pattern fits
together in an intricate way as dimensions are compactified and
decompactified.  Some of the earlier conjectures, such as the $N=4$
conjecture, were incorporated whole, while others were incorporated
in a modified or limited form; we will return to the membrane idea in section
4.2.  

These systematic conjectures made possible many new tests, and
evidence for weak/strong duality in string theory accumulated
rapidly.  This led also to the discovery of various new dynamical
ideas, some of which will be described in part~4. 
There is now an integrated picture
of the strongly coupled dynamics, at least for string theories with
enough supersymmetry, and the methods are being extended to theories
with less supersymmetry.

Figure~3 is a schematic picture of the space of string vacua.
\begin{figure}
\begin{center}
\leavevmode
\epsfbox{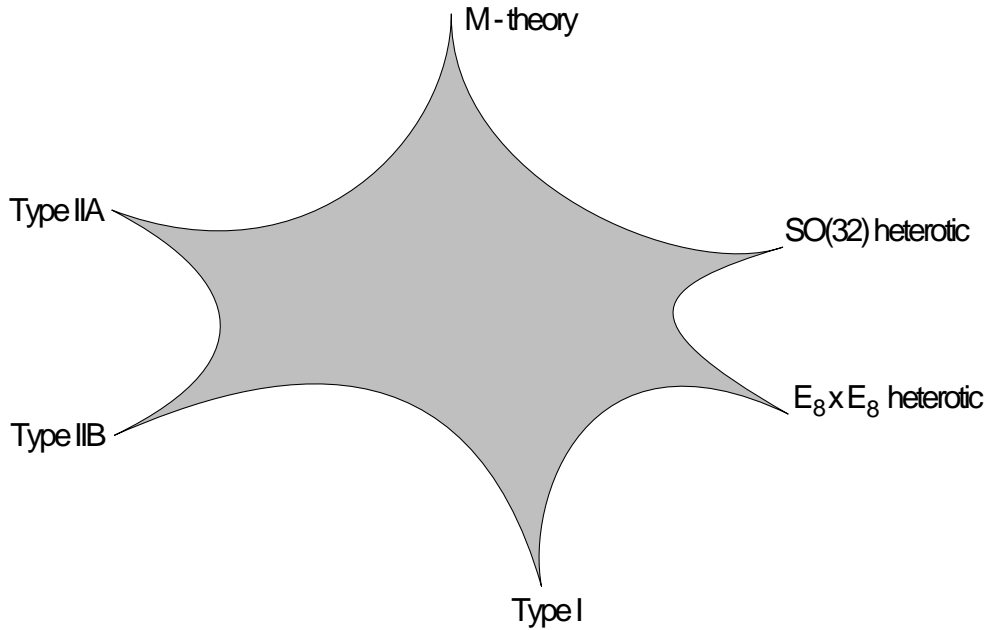}
\caption{Space of string vacua.  The cusps are limits in which a
weakly coupled string description is possible (except for the
M-theory limit).}
\end{center}
\end{figure}
The parameters are the string coupling and the sizes and shapes of
the compact dimensions; in string theory these are not fixed
but are determined by the expectation values of the moduli.
Over most of the space, the string coupling $g$ is of order 1.  In
various limits, there is an effective description in terms of one or
another weakly coupled string theory.  The different string theories
are characterized by the number and kind of supersymmetries 
and by the world-sheet topologies allowed (oriented vs. unoriented and
closed vs. open).  In some limits new theories are encountered; the
most interesting of these, M-theory, will be described in section~4.1.

Figure~3 is actually an oversimplification, in that the different
limits are characterized not only by different string theories but by
different topologies for the compact dimensions.  It has been known
for some time that even in weakly coupled string theory spacetime
topology is not invariant (Aspinwall, Greene, and Morrison, 1993), and
string duality has provided many more examples (Greene, Morrison,
and Strominger, 1995).  Also, a more accurate picture would have many
pieces resembling figure~3, touching one another at points or along
curves.

In field theory the duality multiplets included the elementary quanta 
plus smooth classical configurations (magnetic monopoles).  The string
duality multiplets include these (though the elementary quanta are now
loops rather than points), plus singular classical configurations
(black holes) and also a new type of object unique to
string theory, the D-brane.  D-branes are topological defects on
which the ends of a string can be trapped, as shown in figure~4.
\begin{figure}
\begin{center}
\leavevmode
\epsfbox{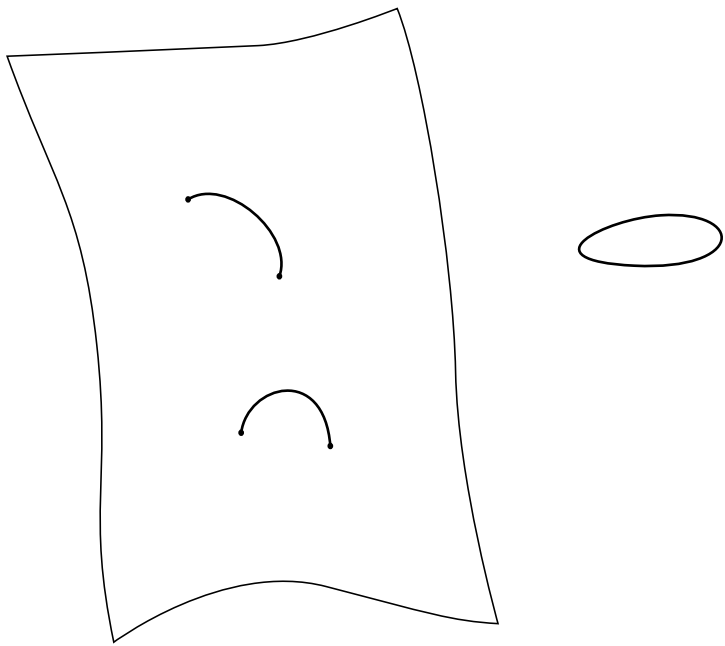}
\caption{D-brane.  Shown are two trapped strings and one not trapped.}
\end{center}
\end{figure}
They
can be pointlike, one-dimensional, two-dimensional, and so on.  These
were discovered in the study of perturbative dualities of string
theories, dualities that relate different weakly coupled theories
(Dai, Leigh, and Polchinski, 1989; Polchinski, Chaudhuri, and
Johnson, 1996).  Their relevance to weak/strong duality was noticed
more recently (Polchinski, 1995).

Let me illustrate one of the methods by which the strongly coupled
limits are determined.  In addition to the fundamental strings,
various string theories have in their spectra one-dimensional objects
which are either smooth solitons or D-branes.  At weak coupling
these are much heavier than the fundamental strings, but at strong
coupling they are much lighter (again this is guaranteed by the BPS
formula).  In this limit it is natural to reinterpret the theory with
the soliton or D-brane being the fundamental string.  Which string
theory one gets depends on the degrees of freedom of the object, which
can be determined by a weak-coupling calculation.  This has been
applied to determine, or confirm, the strongly coupled limits of
various string theories (Sen, 1995a; Harvey and Strominger,
1995; Dabholkar, 1995; Hull,
1995; Schwarz, 1995; Witten, 1996a; Polchinski and Witten, 1996).

\section{Some Results}

I have described the principal methods and general results of string
duality.  In this final part I would like to focus on some 
specific highlights, some of the discoveries that have most greatly
changed our understanding.

\subsection{M-Theory and the Eleventh Dimension}

As described in section~3.4, weakly coupled string theory requires
specifically ten dimensions.  One of the striking consequences of
string duality is the existence of an eleventh dimension, not visible
in weakly coupled string theory.

How does one discover a new dimension?  The experimental signature 
was discussed in section~3.4, a tower of new states above the energy
$1/R$.  The theoretical signature is the same.  A certain string
theory, the IIA theory in ten dimensions, has zero-dimensional
D-branes, D-particles. These are heavy, with a mass $M_{\rm s}/g$ at
weak coupling ($M_{\rm s}$ is the string mass scale).  For a pair of
D-particles, the supersymmetric bound state problem can be solved and
there is a single bound state with mass $2M_{\rm s}/g$ (Witten,
1996a; Sen, 1995b).  Similarly,
$n$ D-particles have a single bound state of mass $nM_{\rm s}/g$.
This bound state spectrum was actually inferred from lower-dimensional
duality symmetries before the explicit construction of the bound
states.  At strong coupling all
of these become light, and in the
$g\to 0$ limit form a continuum.  We have seen just such a spectrum
before, in section~3.4.  It is the signature of a new, eleventh,
dimension with $ R= g/M_{\rm s}$ (Townsend, 1995, Witten,
1995a).  This makes it clear why this dimension is invisible in string
perturbation theory: the latter is an expansion in $g = RM_{\rm s}$,
so an expansion around the zero-radius limit.

Eleven dimensions is an interesting number.  This is the maximum in
which supersymmetry is possible---beyond eleven the massless
multiplets would contain spins higher than two, something which seems
to be impossible in a consistent theory.  Eleven-dimensional
supergravity is therefore the most symmetric theory based on
supersymmetry (Cremmer, Julia, and Scherk, 1978), and was considered
as a possible unifying idea before string theory.  The principal
problems were that it is nonrenormalizable and that it is not possible
to obtain a chiral spectrum.

String theorists generally regarded eleven-dimensional supergravity as
an irrelevant curiosity because of its difficulties and because
strings seemed to live in ten, though some supergravity experts
retained an interest in the theory because of its high degree of
symmetry.  Now it develops that strings have an eleven-dimensional
limit whose low energy physics is determined by symmetry
considerations to be eleven-dimensional supergravity and whose short
distance physics is not understood.  This has been given the
provisional name `M-theory.'

\subsection{Strings from Membranes}

The following exercise gives an interesting insight into M-theory.
Consider a long IIA string in ten dimensions, shown in
figure~5a.
\begin{figure}
\begin{center}
\leavevmode
\epsfbox{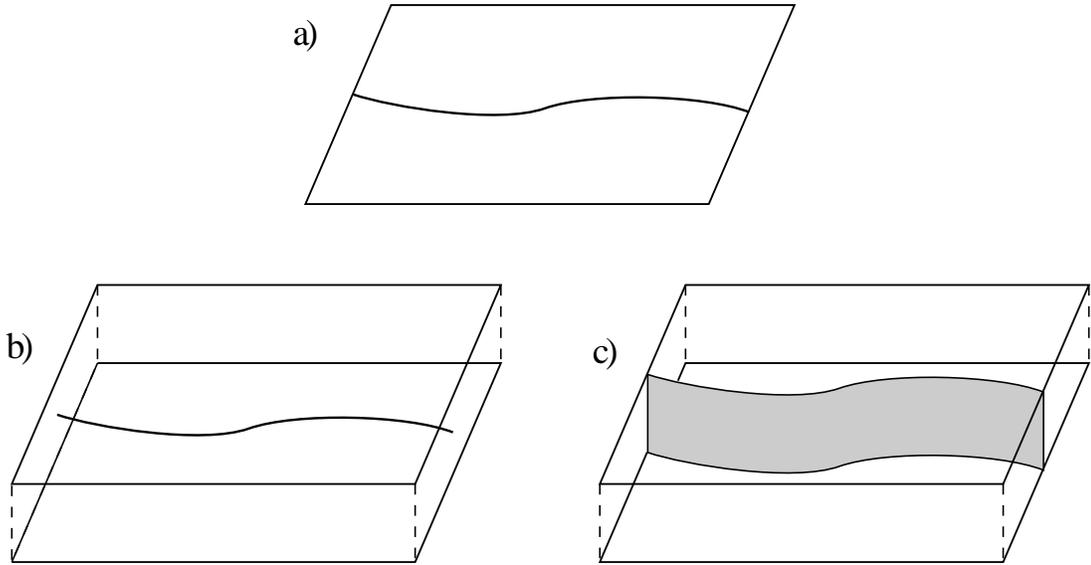}
\caption{a) Long IIA string in ten dimensions.
At strong coupling a perpendicular direction, which is periodic,
becomes visible.  b) String in eleven dimensions.  c) Wrapped membrane
in eleven dimensions.}
\end{center}
\end{figure}
Follow this state as the coupling is increased.  We
now know that a new orthogonal dimension will appear, but there are
two possibilities for the form of the resulting state: in eleven
dimensions it could look like a string as in figure~5b, or it could
turn out to be a  membrane wrapped around the eleventh dimension as
shown in figure~5c.
The answer can be determined in various
ways---from the conserved charges carries by the various objects, from
the scaling of the string tension with coupling, from symmetry---and
it is figure~5c, the wrapped membrane.\footnote
{The supersymmetric membrane was found by Bergshoeff, Sezgin, and
Townsend (1987).  That a wrapped membrane had the same physics as a
ten-dimensional IIA superstring was shown by Duff, Howe,
Inami, and Stelle (1987).}

This is a puzzle.  It seems to say that string theory is really the
quantum mechanics of membranes, but in section~2.1 we have said that
theories of fundamental membranes do not seem to exist.

It is possible that the problems of quantizing membranes will be
overcome and string theory will become membrane theory---this was one of the
origins of the name M-theory.\footnote{M-theory also has
five-dimensional objects, five-branes.}  It is likely, though, that
the resolution lies in a different direction, that strings and
membranes will turn out to be composites of something else.  This has
long been suspected in the case of string theory.  Some of the reasons
for this are reviewed in Polchinski (1994).  They include the general lack of
success of string field theory and the fact that string perturbation
theory diverges more rapidly than in field theory (Shenker, 1991). 
String duality seems to give further support for this.  In figure~3,
the {\it asymptotic} behaviors in various limits are generated by
different string theories, but there is no sign that the string
descriptions do more than this.
Sums over string or membrane world-sheets are directly analogous to
the Feynman graph expansion and so are intrinsically perturbative.
As for duality in field theory, we need a more quantum-mechanical
description, one in which $\hbar$ is not a free parameter (Witten,
1996c).

Incidentally, the membrane can also be oriented perpendicular to the
eleventh dimension, so that in ten dimensions it still looks like a
membrane.  In string theory it appears as a D-brane.  From the
eleven-dimensional point of view one sees a symmetry in which a
$90^\circ$ rotation takes the eleventh dimension into one of the
others.  From the string theory point of view this is a
nonperturbative symmetry, mixing radii and couplings and relating
strings to D-branes.

\subsection{Unification of Couplings}

In this section I would like to explain how all this business of
extra dimensions might enter into an important piece of physics,
the unification of the couplings.  Figure~6a shows the three gauge
couplings and the dimensionless gravitational coupling $G_N E^2$ as a
function of energy.
\begin{figure}
\begin{center}
\leavevmode
\epsfbox{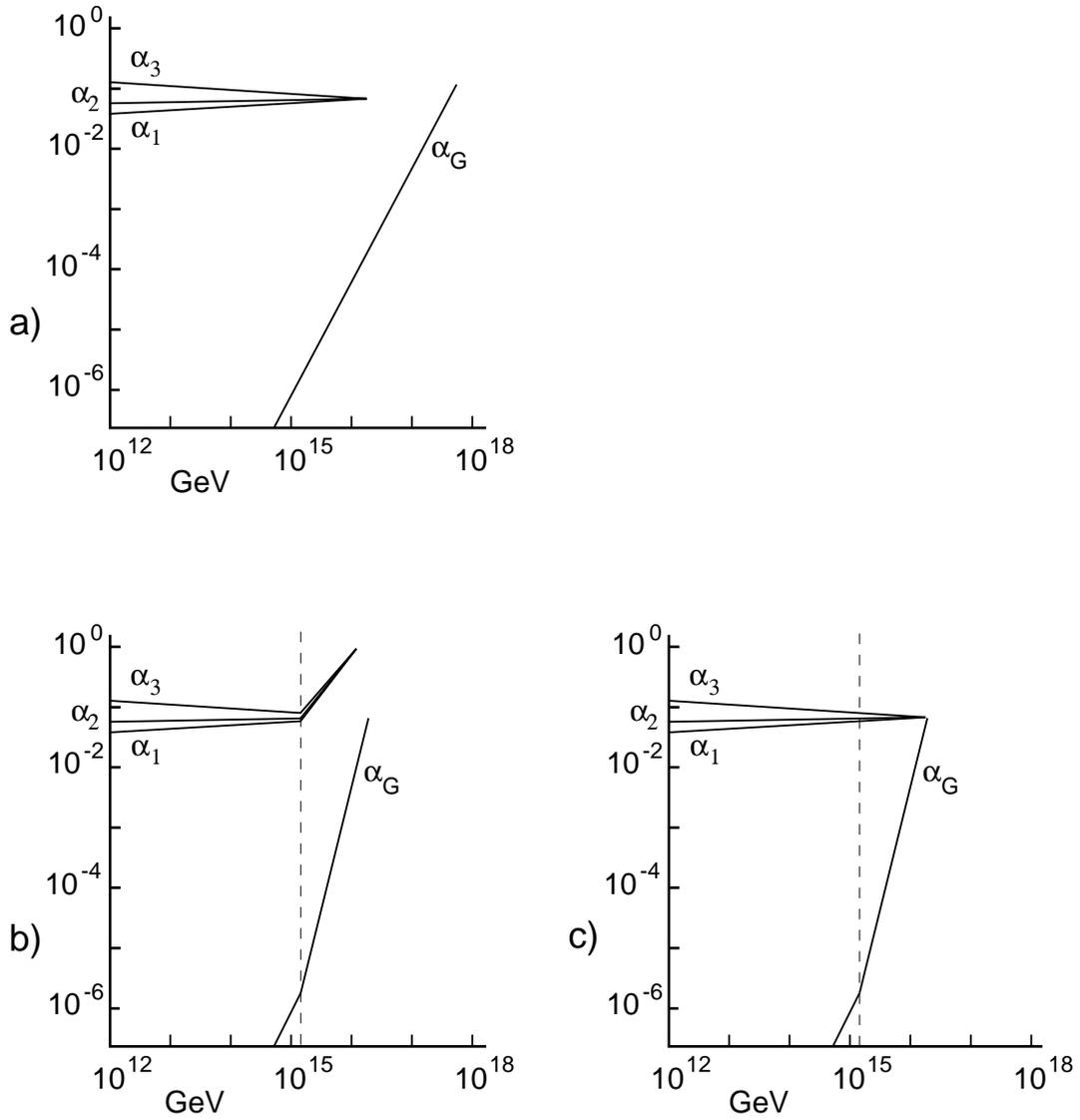}
\caption{a) Running of the three gauge couplings and the dimensionless
gravitational coupling with energy.  b) Effect of a fifth dimension
below the unification scale.  c) Effect of a fifth dimension of
Horava-Witten type.}
\end{center}
\end{figure}
The logarithmic running of the gauge couplings
is familiar (Amaldi, de Boer, and Furstenau, 1991), as is the fact
that they meet within experimental errors in the minimal supersymmetric
model.  It is also striking that the gravitational coupling nearly meets
the other three, missing by a little over one order of magnitude out of
fourteen.\footnote{The unification scale is about three orders of
magnitude below the Planck scale, but accounting for $4\pi$'s and other
factors reduces this to a factor of 20 (Kaplunovsky, 1988).}  This is a
near miss, less than ten percent in scale, but it is larger than the
experimental error and so must be understood.  There are many ways that
the four couplings might turn out to unify.  Additional particles at an
intermediate scale would change the running of the gauge couplings and
can cause them to meet at a higher energy.  Enough extra states near the
unification scale can have the same effect.  Or, the three gauge
couplings may actually unify at the lower energy and a grand unified
field theory with a single gauge coupling describe the physics from
there until the string scale.

All of these ideas modify the behavior of the gauge couplings.  Since
the gauge couplings already meet, it would seem more economical to
find instead a mechanism to change the running of the gravitational
coupling.  But this seems impossible because the running of the
gauge coupling is just dimensional, as $E^2$; gravity is still
classical at these scales.

It is interesting to consider the effect of a new dimension below the
unification scale.  This changes the dimensional analysis.
Both the gauge and gravitational couplings grow more rapidly as shown
in figure~6b, but they do not meet any sooner.

In the $E_8 \times
E_8$ heterotic string, the strong coupling limit leads to a new
dimension which is slightly different from that considered before. 
Instead of being periodic, a circle, it is a segment (Horava and
Witten, 1996).  The gauge fields and matter live at the endpoints only,
while gravity propagates in the bulk.  Suppose that a fifth dimension
of this type exists below the unification scale.  That is, spacetime
is a narrow five-dimensional layer bounded by four-dimensional walls. 
Since the gauge fields and matter live in the walls, the evolution of
the gauge couplings is the same as in four dimensions, but since
gravity propagates in the full five dimensions the gravitational
coupling behaves as in figure~6b. The result is shown in figure~6c:
for a fifth dimension of appropriate size, the kink in the
gravitational coupling makes it meet the others at the unification
scale (Witten, 1996b).

This is no more predictive than the idea of a unified theory: it adds
one new parameter, the size of the fifth dimension.  Nevertheless
it is an economical solution to the problem of the unification of all
couplings, as likely to be true as any other.  With only the data
of the four couplings, there is not enough to distinguish among
the possibilities.  If supersymmetry is found and the masses of the
superpartners measured with precision, many new renormalization group
analyses become possible and it may be possible to determine the physics
at the unification scale.

\subsection{Gauge Invariance}

Local symmetries---gauge invariance and coordinate invariance---are a
key ingredient in physics, appearing in the strong, weak,
electromagnetic, and gravitational interactions.  Yet as has often
been noted, these are not really symmetries in the usual sense. 
Rather, they are {\it redundancies}: different vector potentials or
coordinate systems describe the same physical state.  Why redundancy
should be so important is puzzling, but it seems to be, and most
attempts to unify are based on embedding the local symmetries at low
energy into larger groups such as $SU(3) \times SU(2) \times U(1) \to
SU(5)$.  In weakly coupled four dimensional theories, the size of the
gauge group can only get larger at higher energies.

There have been various arguments made that at strong coupling the
reverse could be true, but convincing examples have only come with
the understanding of strongly coupled dynamics in supersymmetric
gauge and string theories.  There are now many examples of composite
gauge fields in the low energy theory, gauge fields not among the
original short-distance fields (Intriligator and Seiberg, 1995).  The
interpretation of gauge symmetry is reversed: gauge symmetry is
ubiquitous not because it is present in the underlying ultraviolet
theory, but because it is infrared stable.  That is, it is one of the
few kinds of interesting long-distance dynamics which is {\it
natural}, which doesn't disappear when small changes are made in the
short-distance parameters.

\subsection{Black Holes}

In the early 1970's it was found that classical 
black holes obey laws directly analogous to the laws of
thermodynamics.\footnote
{For a review, see for example Carter (1979).} This analogy was made
sharper by Hawking's discovery (1975) that black holes radiate as
black bodies at the corresponding temperature.  Under this analogy,
the entropy of a black hole (the Bekenstein-Hawking entropy) is the
area of its horizon divided by $4 l_{\rm P}^2$.  

It has long been a goal to find a statistical mechanical theory
associated with this thermodynamics, and in particular to associate the
entropy with the density of states of the black hole.  Many arguments
(too many to review here) have been put forward in this direction. 
While it may develop that some of the principles behind these are
correct, until recently there was no example where the
states of a black hole could be counted in a controlled way.

The idea is to count supersymmetric (BPS) black hole states. Such
black holes always carry gauge charges and have the maximum allowed
charge to mass ratio---they are extremal.  As discussed in
section~2.2, the number of BPS states cannot change as we vary the
string coupling constant.  As the coupling is reduced
and so the gravitational interaction weakened, certain black holes
will at sufficiently weak coupling no longer look like black
holes.\footnote{This depends on how the mass scales with the
coupling.  Other black holes continue to look like black holes no
matter how weak the coupling becomes.  The BPS strategy was applied
to these by Larsen and Wilczek (1995), but here one does not have an
explicit understanding of the space of states even at weak coupling.} 
Rather, they look like a collection of D-particles as depicted in figure~7;
this has the same charge-to-mass ratio as the black hole. 
\begin{figure}
\begin{center}
\leavevmode
\epsfbox{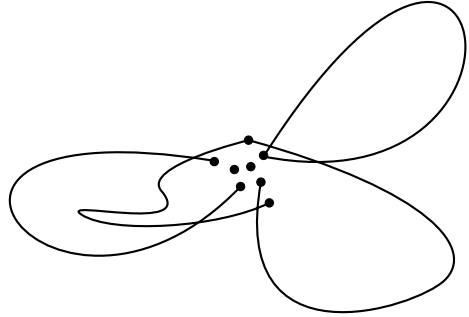}
\caption{Collection of weakly coupled D-particles with some attached
strings, the weakly coupled limit of certain black holes.}
\end{center}
\end{figure}
The D-particles are a weakly coupled quantum
system whose spectrum is explicitly known from D-brane methods.  The
density of BPS states is just that given by the Bekenstein-Hawking
entropy (Strominger and Vafa, 1996, and much subsequent work; for a
review see Horowitz, 1996).

Closely associated with all this is the black hole information
paradox.  A black hole of given mass and charge can be formed in a
very large number of ways.  It will then evaporate, and the final
state is black body radiation that does not depend on how the black
hole formed.  Thus, many initial states evolve to a single final
state.  This violates the usual laws of quantum mechanics.

There are various schools of thought here (reviewed in Page, 1993).  The
proposal of Hawking (1976) is that this is just the way things are: the
laws of quantum mechanics need to be changed.  There is also strong
opposition to this view.  The problem is not that theorists believe
that the laws of quantum mechanics are in their final form---they may
be, though most of us would like to see a less mysterious structure. 
Rather, it is that the specific modification required here, the
replacement of wavefunctions with density matrices, seems ugly and very
possibly inconsistent.

The principal alternative, that the initial state is encoded in
subtle correlations in the Hawking radiation, sounds plausible but
in fact is even more radical.\footnote{The third
major alternative is that the evaporation ends in a remnant, a
Planck-mass object having an enormous number of internal states.  This
might be stable or might release its information over an exceedingly
long time scale.  My own interpretation of string duality is that it
disfavors this idea, as near-extremal black holes have not been found
to have the enormous number of states that it would require.} The problem
is that Hawking radiation emerges from the region of the horizon, where
the geometry is smooth and so ordinary low energy field theory should be
valid.  One can follow the Hawking radiation and see correlations
develop between the fields inside and outside the black hole; the
superposition principle then forbids the necessary correlations to exist
strictly among the fields outside.  To evade this requires that the
locality principle in quantum field theory break down in some
long-ranged but subtle way.  There is a proposal for how this might
occur in string theory (Susskind, 1993, 1995; Lowe, Polchinski,
Susskind, Thorlacius, and Uglum, 1995), but it is controversial.

The recent progress in string duality suggests that black holes do obey
the ordinary rules of quantum mechanics.  The multiplets include black holes along with
various nonsingular states, and in figure~7 we have continuously
deformed a black hole into a system which obeys ordinary quantum
mechanics, at least to high accuracy.  This is certainly not decisive,
however.  The problem is that dynamical properties are not like the
counting of BPS states; they are not invariant under changes in
parameters. It could be that
as the coupling constant is increased, a critical coupling is reached
where the D-particles collapse into a black hole.  At this coupling
there could be a discontinuous change (or a smooth crossover) from
ordinary quantum behavior to information loss.
It is an open question whether D-branes will give
insight into the dynamical behavior of black holes; there are hints
that it may be so.

\section{Conclusions}

So when will string theory make sharp predictions?  Probably not until
the vacuum is understood, as we discussed in section~2.3.  An
important problem remains in our understanding of the vacuum, when
we try to calculate the vacuum energy density, the cosmological
constant.  The vacuum is a complicated place, and there are many
different contributions to the energy density---the zero point
energies of the quantum fields, the potential energy of the Higgs
field, the energy of the strongly fluctuating QCD fields.  The total
of the separate densities is at least $10^8$ GeV$^4$, and there
is no reason known that they should cancel to any degree of accuracy. 
Yet the experimental bound on the cosmological
constant is 55 orders of magnitude smaller than this (see Weinberg,
1989, for a review).

This is just as much problem in quantum field theory as in string
theory but field theory can still make precise predictions because
it is less ambitious.  One can ignore gravity, or add a free
parameter and adjust it to cancel the other contributions to the
cosmological constant.  In string theory, neither of these is
possible.

Does string duality help here?  Possibly.  It is an interesting fact
that in supersymmetric theories the energy density can cancel
naturally.  For example the zero point energies of bosons and their
fermionic partners are equal and opposite.  One can see this in the
discussion of BPS states: a supersymmetric vacuum has zero energy
(though the inclusion of supergravity makes things a bit more subtle).
In nature, though, supersymmetry must be spontaneously
broken---there is no charged boson degenerate with the electron---and
spontaneous breaking spoils the cancellation of the cosmological
constant.  We need some new phase of supersymmetric theories, in which
the boson/fermion degeneracy is removed while the vacuum energy
remains zero.  Witten (1995b) has suggested how such a phase might
appear in strongly coupled string theory.

From another point of view, the cosmological constant and
black hole information problems both seem to require subtle
nonlocalities in spacetime.  For the information problem we have
discussed this; for the cosmological constant, the point is that the
low energy value of this constant must somehow feed back into the
short-distance physics that determines it.  For example, it was
argued that spacetime wormholes could accomplish this (Coleman,
1988).  This proposal was the subject of intense and skeptical
scrutiny, but the subject ground to a halt because a nonperturbative
formulation of quantum gravity was lacking. It may then be that it
will return, possibly in some transmuted form, when we know what
string theory is.

In closing, let me say that there is a sense among those working in
string theory that we are dealing with a unique and remarkable
structure, one which has many points of contact with the physics
we know and with earlier attempts to unify it---quantum mechanics,
gravity, gauge symmetry, chirality, grand unification,
supersymmetry, and Kaluza-Klein theory.  There is also a sense that
we are {discovering} this structure, not inventing it.

The
distinction is illustrated, for example, by the question of the
nature of spacetime.  There have been many previous attempts to
modify spacetime in a way that would look the same at long distance
but would solve the problems of quantum gravity.  This approach has
often been sterile---it is not a problem that seems to yield to direct
effort.  String theory, on the other hand, starts with a flat
spacetime.  One discovers first that the spacetime is dynamical, that
the theory contains gravity.  Later one finds, as mentioned in
section~3.5, that spacetime topology is not a physical invariant
but can change in specific and controlled ways.  There is still more
to be learned.  In perturbation theory it seems that the shortest
sensible distance is the string length scale, the
typical radius of zero point vibrations of the string 
(Gross and Mende, 1987; Amati, Ciafaloni, and Veneziano, 1989;
reviewed in Witten, 1996c). This works out
to be the Planck length divided by a power of the string coupling.
There is evidence for a somewhat shorter scale in the exact theory
(Shenker, 1995), and that D-branes may probe it (Bachas, 1995;
Danielsson, Ferretti and Sundborg, 1995; Kabat
and Pouliot, 1995).  Moreover, there is a sense
in which the spacetime coordinates for D-branes are elevated from
numbers to matrices (Witten, 1996a); only at low energy the matrices
are diagonal and an ordinary spacetime picture holds.  It may turn
that this is a curiosity, or it may signal a new uncertainty
principle relating to a minimum distance.
Exploring the connection between D-branes and black holes is
a likely way to learn which.  In effect, string theory is smarter than
we are.  It knows what spacetime is, and we don't, and we have to figure
out how to ask it.

We are very fortunate that this remarkable structure exists and that
it seems to be within our power to understand it.

\subsection*{Acknowledgements}

I would like to thank Gary Horowitz, Clifford Johnson, and Andy
Strominger for their suggestions on the manuscript.  This work is
supported by NSF grants PHY91-16964 and PHY94-07194.

\newpage
\section*{References}
\baselineskip=16pt
U. Amaldi, W. de Boer, and H. Furstenau, 1991,
``Comparison of Grand Unified Theories with Electroweak and Strong
Coupling Constants Measured at LEP,'' Phys. Lett. {\bf B260,} 447.
\\[4pt]
D. Amati, M. Ciafaloni, and G. Veneziano, 1989, ``Can Spacetime Be
Probed Below the String Scale?'' Phys. Lett. {\bf B216,} 41.
\\[4pt]
P. S. Aspinwall, B. R. Greene, and D. R. Morrison, 1993, ``Multiple
Mirror Manifolds and Topology Change in String Theory,''
Phys. Lett. {\bf B303,} 249, e-print hep-th/9301043.
\\[4pt]
C. Bachas, 1996, ``D-Brane Dynamics,'' Phys. Lett.
{\bf B374,} 37, e-print hep-th/9511043.
\\[4pt]
E. Bergshoeff, E. Sezgin, and P.K. Townsend, 1987,
``Supermembranes and Eleven Dimensional Supergravity,''
Phys. Lett. {\bf B189,} 75.
\\[4pt]
P. Candelas, G. Horowitz, A. Strominger, and E. Witten, 1985,
``Vacuum Configurations for Superstrings,'' Nucl. Phys. {\bf B258,}
46.
\\[4pt]
B. Carter, 1979, ``The General Theory of the Mechanical,
Electromagnetic, and Thermodynamic Properties of Black Holes,''
in {\it General Relativity. An Einstein Centennial Survey,}
ed. S. W. Hawking and W. Israel (Cambridge, Cambridge).
\\[4pt]
S. Coleman, 1988, ``Why There is Nothing Rather Than Something: A
Theory of the Cosmological Constant,'' 
Nucl. Phys. {\bf B310,} 643.
\\[4pt]
E. D. Commins and P. H. Bucksbaum, 1983, {\it Weak Interactions of
Leptons and Quarks} (Cambridge, New York).
\\[4pt]
E. Cremmer, B. Julia, and J. Scherk, 1978, ``Supergravity Theory in
Eleven Dimensions,''  Phys. Lett. {\bf 76B,} 409.
\\[4pt]
A. Dabholkar, 1995, ``Ten-Dimensional Heterotic String as a
Soliton,'' Phys. Lett. {\bf B357,} 307, e-print hep-th/9506160.
\\[4pt]
J. Dai, R. G. Leigh, and J. Polchinski, 1989, ``New Connections Between
String Theories,'' Mod. Phys. Lett. {\bf A4,} 2073. 
\\[4pt]
U. H. Danielsson, G. Ferretti, and B. Sundborg, 1996,
``D-Particle Dynamics and Bound States,''
preprint USITP-96-03, e-print hep-th/9603081. 
\\[4pt]
P. A. M. Dirac, 1931, ``Quantized Singularities in the
Electromagnetic Field,'' Proc. Roy. Soc. {\bf A133,} 60.
\\[4pt]
M. J. Duff, P.S. Howe, T. Inami, and K.S. Stelle, 1987,
``Superstrings in $D=10$ from Supermembranes in $D=11$,''
Phys. Lett. {\bf 191B,} 70. 
\\[4pt]
M. J. Duff, 1988, ``Supermembranes: The First Fifteen Weeks,''
Class. Quant. Grav. {\bf 5,} 189.
\\[4pt]
M. J. Duff, 1995, ``Strong/Weak Duality from the Dual String,''
Nucl. Phys. {\bf B442,} 47, e-print hep-th/9506057.
\\[4pt]
%M. J. Duff, S. Ferrara, R. R. Khuri, and J. Rahmfeld, 1995,
%``Supersymmetry and Dual String Solitons,''
%Phys. Lett. {\bf B356,} 479, e-print hep-th/9506057. 
%\\[4pt]
A. Font, L. Ibanez, D. Lust, and F. Quevedo, 1990, ``Strong-Weak
Coupling Duality and Nonperturbative Effects in String Theory,''
Phys. Lett. {\bf B249,} 35.
\\[4pt]
F. Gliozzi, J. Scherk, and D. Olive, 1977, ``Supersymmetry,
Supergravity Theories, and the Dual Spinor Model,''
Nucl. Phys. {\bf B122,} 253; reprinted in Schwarz, 1985, Vol. 1, 300.
\\[4pt]
M. B. Green and J. H. Schwarz, 1984, ``Anomaly Cancellations in
Supersymmetric $D=10$ Gauge Theory and Superstring Theory,''
Phys. Lett. {\bf 136B,} 367; reprinted in Schwarz, 1985, Vol. 2, 1007.
\\[4pt]
M. B. Green, J. H.
Schwarz, and E. Witten, 1987, {\it Superstring Theory, Vols. 1 and 2}
(Cambridge, Cambridge).
\\[4pt]
B. R. Greene, D. R. Morrison, and
A. Strominger, 1995, ``Black Hole Condensation and the Unification of
String Vacua,'' Nucl. Phys. {\bf B451,} 109, e-print hep-th/9504145 .
\\[4pt]
D. J. Gross, J. A. Harvey, E. Martinec, and R. Rohm, 1985, ``Heterotic
String,'' Phys. Rev. Lett. {\bf 54,} 502.
\\[4pt]
R. Haag, J. T. Lopuszanski, and M Sohnius, 1975,
``All Possible Generators of Supersymmetries of the S Matrix,''
Nucl. Phys. {\bf B88,} 257. 
\\[4pt]
J. A. Harvey and A. Strominger, 1995, ``The Heterotic String is a
Soliton,'' Nucl. Phys. {\bf B449,} 535, (E) {\bf B458,} 456, 
e-print hep-th/9504047.
\\[4pt]
J. A. Harvey, 1996, ``Magnetic Monopoles, Duality, and
Supersymmetry,'' Chicago preprint EFI-96-06, e-print hep-th/9603086.
\\[4pt]
S. W. Hawking, 1975, ``Particle Creation by Black Holes,''
Comm. Math. Phys. {\bf 43,} 199.
\\[4pt]
S. W. Hawking, 1976, ``Breakdown of Predictability in Gravitational
Collapse,'' Phys. Rev. {\bf D14,} 2460.
\\[4pt]
P. Horava and E. Witten, 1996, ``Heterotic and Type I String Dynamics
from Eleven Dimensions,'' Nucl. Phys. {\bf B460,} 506, 
e-print hep-th/9510209.
\\[4pt]
G. T. Horowitz, 1996, ``The Origin of Black Hole Entropy in String
Theory,'' UCSB preprint UCSBTH-96-07, e-print gr-qc/9604051.
\\[4pt]
C. M. Hull and P. K. Townsend, 1995, ``Unity of Superstring
Dualities,'' Nucl. Phys. {\bf B438,} 109, e-print hep-th/9410167. 
\\[4pt]
C. M. Hull, 1995, ``String-String Duality in Ten-Dimensions,''
Phys. Lett. {\bf B357,} 545, 
e-print hep-th/9506194.
\\[4pt]
K. Intriligator and N. Seiberg, 1995,
``Lectures on Supersymmetric Gauge Theories and Electric - Magnetic
Duality,'' presented at TASI 95, Boulder, Colo., 
Trieste Summer School, Trieste, Italy, and
Cargese Summer School, Cargese, France, Rutgers preprint RU-95-48,
e-print hep-th/9509066.
\\[4pt]
D. Kabat and P. Pouliot, ``A Comment on Zero-Brane Quantum
Mechanics,'' Rutgers preprint RU-96-17, e-print hep-th/9603127.
\\[4pt]
V. S. Kaplunovsky, 1988, ``One Loop Effects in String Unification,''
Nucl. Phys. {\bf B307,} 145, (E) {\bf B382,} 436 (1992), 
e-print hep-th/9205070.
\\[4pt]
D. A. Lowe, J. Polchinski, L. Susskind, L. Thorlacius, and
J. Uglum, 1995, ``Black Hole Complementarity versus
Locality,'' Phys. Rev. {\bf D52,} 6997, e-print hep-th/9506138.
\\[4pt]
C. Montonen and D. Olive, 1977, ``Magnetic Monopoles as Gauge
Particles?'' Phys. Lett. {\bf 72B,} 117.
\\[4pt]
A. Neveu and J. Scherk, 1972, ``Connection Between Yang-Mills Fields
and Dual Models,'' Nucl. Phys. {\bf B36,} 155.
\\[4pt]
H. Osborn, 1979, ``Topological Charges for $N=4$ Supersymmetric Gauge
Theories and Monopoles of Spin 1,'' Phys. Lett. {\bf 83B,} 321. 
\\[4pt]
D. N. Page, 1993, ``Black Hole Information,''
Canadian Gen. Rel. {\bf 1993,} 1, e-print hep-th/9305040.
\\[4pt]
J. Polchinski, 1994, ``What is String Theory?'' 
Les Houches Summer School,
Session 62: Fluctuating Geometries in Statistical Mechanics and Field
Theory, 1994, e-print hep-th/9411028
\\[4pt]
J. Polchinski, 1995, ``Dirichlet-Branes and Ramond-Ramond Charges,''
Phys. Rev. Lett. {\bf 75,} 4724, e-print hep-th/9510017.
\\[4pt]
J. Polchinski and E. Witten, 1996, ``Evidence for Heterotic - Type I
String Duality,'' Nucl. Phys. {\bf B460,} 525, e-print hep-th/9510169.
\\[4pt]
J. Polchinski, S. Chaudhuri, and C. V. Johnson, 1996, ``Notes on
D-Branes,'' ITP preprint NSF-ITP-96-003, e-print hep-th/9602052.
\\[4pt]
P. Ramond, 1971, ``Dual Theory for Free Fermions,'' Phys. Rev.
{\bf D3,} 2415; reprinted in Schwarz, 1985, Vol. 1, 53.
\\[4pt]
A. Salam, 1980, ``Grand Unification of Fundamental Forces,'' Rev. Mod.
Phys. {\bf 52,} 525.
\\[4pt]
J. Scherk and J. H. Schwarz, 1974, ``Dual Models for
Non-Hadrons,'' Nucl. Phys. {\bf B81,} 118;
reprinted in Schwarz, 1985, Vol. 1, 191.
\\[4pt]
J. Scherk and J. H. Schwarz, 1975, ``Dual Model Approach to a
Renormalizable Theory of Gravitation,'' Caltech preprint
CALT-58-488; reprinted in Schwarz, 1985, Vol. 1, 218.
\\[4pt] 
J. H. Schwarz, 1985, {\it Superstrings.  The First 15
Years of Superstring Theory, Vols. 1 and 2} (World Scientific,
Singapore).
\\[4pt]
J. H. Schwarz, 1995, ``An $SL(2,Z)$ Multiplet of Type IIB
Superstrings,'' Phys. Lett. {\bf B360,} 13, (E) {\bf B364,} 252,
e-print hep-th/9508143.
\\[4pt]
N. Seiberg, 1994, ``Exact Results on the Space of Vacua of
Four-Dimensional SUSY Gauge Theories,'' Phys. Rev.
{\bf D49,} 6857, e-print hep-th/9402044.
\\[4pt]
N. Seiberg, 1995, ``The Power of Duality: Exact Results in 4-D SUSY
Field Theory,'' Rutgers preprint
RU-95-37, e-print hep-th/9506077.
\\[4pt]
A. Sen, 1994, ``Strong - Weak Coupling Duality in Four-Dimensional
String Theory,'' Int. J. Mod. Phys. {\bf A9,} 3707, e-print
hep-th/9402002.
\\[4pt]
A. Sen, 1995a, ``String String Duality Conjecture in Six-Dimensions
and Charged Solitonic Strings,''
Nucl. Phys. {\bf B450,} 103, e-print hep-th/9504027.
\\[4pt]
A. Sen, 1995b, ``A Note on Marginally Stable Bound States in Type II
String Theory,'' Mehta preprint MRI-PHY-23-95, e-print hep-th/9510229.
\\[4pt]
S. H. Shenker, 1991, ``The Strength of Nonperturbative Effects in
String Theory,'' in {\it Random Surfaces and Quantum Gravity}, eds. O.
Alvarez, E. Marinari, and P. Windey (Plenum, 1991).
\\[4pt]
S. H. Shenker, 1995, ``Another Length Scale in String Theory?''
Rutgers preprint RU-95-53, e-print hep-th/9509132.
\\[4pt]
A. Strominger, 1990, ``Heterotic
Solitons,'' Nucl. Phys. {\bf B343,} 167, (E) {\bf B353,} 565.
\\[4pt]
A. Strominger and C. Vafa, 1996, ``Microscopic Origin of the
Bekenstein-Hawking Entropy,'' preprint HUTP-96-A002, e-print
hep-th/9601029.
\\[4pt]
L. Susskind, 1993, ``String Theory and the Principles of Black Hole
Complementarity,''  Phys. Rev. Lett. {\bf 71,} 2367,
e-print hep-th/9307168 
\\[4pt]
L. Susskind, 1995, ``The World as a Hologram,''
J. Math. Phys. {\bf 36,} 6377, e-print hep-th/9409089. 
\\[4pt]
P. K. Townsend, 1995, ``The Eleven-Dimensional Supermembrane
Revisited,'' Phys. Lett. {\bf B350,} 184, e-print hep-th/9501068.
\\[4pt]
S. Weinberg, 1980, ``Conceptual Foundations of the Unified
Theory of Weak and Electromagnetic Interactions,'' Rev. Mod. Phys.
{\bf 52,} 515.
\\[4pt]
S. Weinberg, 1989, ``The Cosmological Constant Problem,''
Rev. Mod. Phys. {\bf 61,} 1. 
\\[4pt]
K. Wilson, 1983, ``The Renormalization Group and Critical Phenomena,''
Rev. Mod. Phys. {\bf 55,} 515.
\\[4pt]
E. Witten and D. Olive, 1978, ``Supersymmetry Algebras that Include
Topological Charges,'' Phys. Lett. {\bf 78B,} 97.
\\[4pt]
E. Witten, 1981, ``Search for a Realistic Kaluza-Klein Theory,''
Nucl. Phys. {\bf B186,} 412.
\\[4pt]
E. Witten, 1995a, ``String Theory Dynamics in Various Dimensions,''
Nucl. Phys. {\bf B443,} 85, e-print hep-th/9503124.
\\[4pt]
E. Witten, 1995b, ``Strong Coupling and the Cosmological Constant,''
Mod. Phys. Lett. {\bf A10,} 2153, e-print hep-th/9506101.
\\[4pt]
E. Witten, 1996a, ``Bound States of Strings and $p$-Branes,''
Nucl. Phys. {\bf B460,} 335, e-print hep-th/9510135. 
\\[4pt]
E. Witten, 1996b, ``Strong-Coupling Expansion of Calabi-Yau
Compactification,'' IAS preprint IASSNS-HEP-96-08,
e-print hep-th/9602070.
\\[4pt]
E. Witten, 1996c, ``Reflections on the Fate of Spacetime,'' Phys.
Today {\bf 49,} 4, 5.
\end{document}